\def\lesssim{\mathrel{\hbox{\rlap{\hbox{\lower4pt\hbox{$\sim$}}}\hbox{$<$}}}}

\def\gtrsim{\mathrel{\hbox{\rlap{\hbox{\lower4pt\hbox{$\sim$}}}\hbox{$>$}}}}

\def\msun{M$_{\odot}$}

\def\ll_lsun{log$({L/\rm L_{\odot}})$~}

\def\masa_msun{$M/ \rm M_{\odot}$~}

\def\m_mstar{$M/M_{*}$~}

\documentclass{aa}

\usepackage{graphicx}

\begin{document}

\title{New nonadiabatic pulsation computations on full  PG1159 
evolutionary models: the theoretical GW Vir instability strip revisited}

\author{A. H. C\'orsico$^{1,2}$\thanks{Member of the Carrera del Investigador
Cient\'{\i}fico y Tecnol\'ogico, CONICET, Argentina.},
L. G. Althaus$^{1,2\star}$
\and M. M. Miller Bertolami$^{1,2,3}$\thanks{Fellow of CONICET, Argentina.}}

\offprints{A. H. C\'orsico}

\institute{
$^1$   Facultad   de   Ciencias  Astron\'omicas   y   Geof\'{\i}sicas,
Universidad  Nacional de  La Plata,  Paseo del  Bosque S/N,  (1900) La
Plata, Argentina.\\ $^2$ Instituto de Astrof\'{\i}sica La Plata, IALP,
CONICET-UNLP\\  $^3$ Max-Planck-Institut f\"ur  Astrophysik, Garching,
Germany\\
\email{acorsico,althaus,mmiller@fcaglp.unlp.edu.ar} }
  
\date{Received; accepted}

\abstract{}{We reexamine  the theoretical 
instability domain of pulsating PG1159 stars (GW Vir variables).}  {
We performed an extensive  $g$-mode stability analysis on  PG1159  evolutionary
models   with  stellar   masses   ranging  from   $0.530$  to   $0.741
M_{\odot}$, for which the complete  evolutionary stages  of  their 
progenitors from the ZAMS, through the thermally pulsing AGB and
born-again phases  to the domain of the  PG1159 stars have been
considered.} {We  found that pulsations  in PG1159 stars  are 
excited by the $\kappa$-mechanism due to partial ionization
of carbon  and oxygen,  and that no  composition gradients  are needed
between the surface  layers and the driving region,  much in agreement
with previous studies. We show, for the first time, the existence  of 
a red edge of the instability strip at high luminosities. 
We found that all  of the GW Vir  stars 
lay within our theoretical  instability strip. 
Our results suggest a qualitative  good agreement between the 
observed and the predicted ranges of unstable periods of 
individual stars. 
Finally, we found that generally the seismic masses 
(derived from the period spacing) of GW Vir stars 
are somewhat different from the masses suggested by evolutionary 
tracks coupled with spectroscopy. Improvements in the 
evolution during the thermally pulsing AGB phase 
and/or during the core  helium burning stage and early  AGB  could help 
to alleviate the persisting discrepancies.}{}


\authorrunning{C\'orsico et al.}

\titlerunning{The theoretical GW Vir instability strip}

\maketitle

 
\section{Background}

Pulsating PG1159 stars  --- after the prototype of  the spectral class
and  the  variable  type,  PG 1159-035  or  GW Vir  ---  are  very  hot
hydrogen-deficient  post-Asymptotic  Giant  Branch  (AGB)  stars  with
surface layers rich  in helium ($\sim 30-85 \%$),  carbon ($\sim 15-60
\%$) and oxygen ($\sim 2-20  \%$) (Werner \& Herwig 2006) that exhibit
multiperiodic,  low degree  ($\ell  \leq 2$),  high  radial order  ($k
\gtrsim 18$) $g$-mode luminosity  variations with periods in the range
from  about  300  to  3000  seconds.  Some GW Vir stars are still embedded in
a planetary nebula; they are commonly called PNNVs (Planetary Nebula 
Nucleus Variable). PNNV stars are characterized by much higher luminosity 
than the ``naked'' GW Vir stars (those without nebulae)\footnote{In this 
paper,  naked GW  Vir stars and  PNNV stars will  be indistinctly
referred  to as  GW  Vir stars  or  pulsating PG1159.}. 
GW Vir stars are
particularly important to infer fundamental properties about pre-white
dwarfs  in   general,  such  as   the  stellar  mass and the surface 
compositional stratification\footnote{In  addition,  pulsating PG1159  
stars  have  recently  been shown  by C\'orsico  \& Althaus  (2005) to  
be valuable  tools to  
constrain the occurrence of extra mixing episodes in their progenitor stars.}
(Kawaler  \& Bradley 1994; C\'orsico \& Althaus 2006).

PG1159  stars are believed to be the   evolutionary connection  
between post-AGB stars and most of the hydrogen-deficient white dwarfs. 
These stars are
thought to be the result of a born again episode triggered either by a
very late helium  thermal pulse (VLTP) occurring in  a hot white dwarf
shortly after  hydrogen burning has almost ceased  (see Fujimoto 1977;
Sch\"onberner 1979 and  more recently Althaus et al.   2005) or a late
helium  thermal  pulse (LTP)  that  takes  place  during the  post-AGB
evolution when  hydrogen burning is  still active (see  Bl\"ocker 2001
for  references). During  a VLTP  episode, most  of  the hydrogen-rich
envelope of  the star  is burnt in  the helium-flash  convection zone,
whilst  in a  LTP hydrogen-deficient  composition is  the result  of a
dilution episode. In both cases,  the star returns rapidly back to the
AGB and finally into the  domain of high effective temperature as a 
hydrogen-deficient, quiescent helium-burning object.

A  longstanding  problem associated  with  pulsating  PG1159 stars  is
related  to  the  excitation  mechanism. The early work by 
Starrfield et  al.  (1983) was successful  in finding
the correct destabilizing agent, namely the  
$\kappa$-mechanism   associated  with  the  partial
ionization of  the K-shell  electrons of carbon  and/or oxygen  in the
envelope of  models. However, their  models required a driving region very
poor  in  helium  in order  to  be  capable to  excite
pulsations; even  very low  amounts of  helium  
could weaken  or completely remove the destabilizing  effect of carbon
and oxygen (i.e. ``helium poisoning'').  
The latter requirement led to the conjecture that
a   composition  gradient   would   exist  to   make  compatible   the
helium-devoid  driving   regions  and  the   helium-rich  photospheric
composition. Even modern detailed calculations still point out the necessity 
of a compositional gradient in the envelopes of models
(Bradley \& Dziembowski 1996; Cox 2003). The  presence  
of a chemical composition gradient  is difficult  to explain 
in  view of  the fact that  PG1159 stars  are still  
experiencing mass  loss  ($\dot{M} \sim
10^{-8.1} M_{\odot}$ yr$^{-1}$ for PG 1159-035; Koesterke et al. 1998),
a fact  that prevents the  action of gravitational settling  of carbon
and oxygen, and instead, tends to homogenize the envelope of hot 
white dwarfs (Unglaub \& Bues 2000). 

Clearly at odds  with the hypothesis of a  composition gradient in the
PG1159 envelopes,  calculations  by  Saio (1996), Gautschy
(1997), and Quirion et al.  (2004) ---  based on  
modern opacity  OPAL data  --- demonstrated 
that $g$-mode pulsations in the correct ranges of effective temperatures 
and periods could be easily excited in PG1159 models 
having an uniform envelope composition. 
The most recent study about PG1159-type pulsations is that of Gautschy et
al.  (2005) (hereinafter  GAS05)  based  on a   full
PG1159 evolutionary sequence started from the zero-age main sequence
(ZAMS) and evolved through the  thermally pulsing and VLTP  phases 
(see Althaus et al. 2005). These authors found no need 
for invoking composition gradients in the  PG1159 envelopes
to promote instability. 

As important as they are, the vast majority of the  studies of pulsation 
driving in PG1159 stars rely on simplified  stellar models.  
Indeed, the  earliest works
employed static envelope models and  old opacity data.  Even more modern
works, although  based on  updated opacity data  (OPAL), still  use a
series of static envelope models that no represent a real evolutionary
sequence,   or   evolutionary   computations   based   on   simplified
descriptions of the evolution of their progenitors.  The
only exception is the  work of GAS05, which employs equilibrium
PG1159  models that evolved  through the AGB and born-again stages,  
beginning  from a  $2.7 M_{\odot}$  zero age  main sequence  model star.  
GAS05  analyzed four model sequences, 
with $0.530,  0.55, 0.589$ and $0.64 M_{\odot}$, being
the $0.589 M_{\odot}$ sequence  derived directly from the evolutionary
computations of  Althaus et al.  (2005). The  remainder sequences were
created from  the $0.589 M_{\odot}$ one by  appropriately changing the
stellar mass  shortly after  the end of  the born-again  episode. 

On the basis of {\it full evolutionary} PG1159 models covering the 
whole range of observed GW Vir masses, this paper is intended to 
confirm and extend the results already put forward by the stability analysis 
by Saio (1996), Gautschy (1997), Quirion et al.  (2004), 
and GAS05. We  analyze the pulsational stability  of seven 
different evolutionary 
sequences of  PG1159 models with stellar masses between $0.530$  
and $0.741 M_ {\odot}$. Here,  all of the PG1159  
evolutionary sequences have been  derived  by considering  
the  complete  evolution  of  their
progenitors, an aspect that constitutes an improvement over previous 
studies.   One  of  such  sequences ($0.589  M_{\odot}$)  is  that
already presented by Althaus et al.  (2005) and analyzed by GAS05, and
the  remaining ones are  those  computed recently  by  Miller Bertolami  \&
Althaus (2006), with the exception of 
the $0.741 M_{\odot}$ sequence, which is 
presented for the first time in this work. 
We believe that the pulsational results presented
here based  on extensive full  evolutionary models shed new  lights on
the  GW  Vir  stars and  place  previous  studies  on a  solid  basis,
regarding stellar modeling. The paper is organized as follow: in 
the next Section we briefly describe the input physics of  
our evolutionary code and the PG1159 evolutionary sequences 
analyzed. A brief description of our nonadiabatic 
pulsation code is presented as well. In 
\S \ref{s-calculations} we elaborate on a detailed description of our 
stability analysis and in \S \ref{theory-observations} 
we compare our predictions with the observed properties of known 
GW Vir stars. Finally, in Sect. \ref{conclusions} we summarize 
our main results and make some concluding remarks.

\section{Evolutionary sequences}
\label{evolutionary}

The nonadiabatic pulsational analysis presented in this work relies on
stellar models  that take into  account the complete evolution  of the
PG1159 progenitor stars. The evolution of  such models has been computed with
the LPCODE  evolutionary code,  which is described  in Althaus  et al.
(2005).   LPCODE uses  OPAL
radiative opacities (including  carbon- and oxygen-rich mixtures) from
the  compilation of  Iglesias \&  Rogers (1996),  complemented  at the
low-temperature  regime  with  the  molecular opacities of Alexander  \&
Ferguson  (1994)  (with  solar  metallicity).   Chemical  changes  are
performed  via  a  time-dependent  scheme that  simultaneously  treats
nuclear evolution and mixing  processes due to convection, salt finger
and   overshooting.    Convective  overshooting   is   treated  as  an
exponentially   decaying  diffusive  process   above  and   below  any
convective region.

Specifically, the background of stellar models has been extracted  from 
the evolutionary  calculations recently presented
in  Miller Bertolami \& Althaus  (2006) and Althaus et al. (2005), who  
computed  the full
evolution of initially 1, 2.2, 2.7, 3.05, and  3.5 \msun. In addition,
we include a new sequence of initially 3.75 \msun.   
All of  the sequences  were evolved from  the ZAMS
through the thermally pulsing and  loss mass phases on the AGB.  After
experiencing several  thermal pulses, the progenitors  depart from the
AGB and evolve towards  high effective temperatures.  Mass loss during
the departure  from the  AGB has been  arbitrarily fixed, 
as  to obtain  a final  helium  shell flash
during  the early  white dwarf  cooling phase.   After  the born-again
episode, the hydrogen-deficient, quiescent helium-burning remnants
evolve  at constant  luminosity to  the  domain of  PG1159 stars  with
surface chemical  composition rich in  helium, carbon and  oxygen. The
masses of  the remnants span the range $0.530-0.741  M_{\odot}$.  
For the sequence of $1 M_{\odot}$ two 
different AGB evolution have been considered, with different 
mass loss rates as to obtain different number of thermal pulses 
and, eventually, two different remnant masses of 0.530 and 
0.542 $M_{\odot}$. The main  characteristics of  the sequences  
considered in  this  work are given in  Table \ref{table2}.  We  list 
the initial and  final stellar
mass  (at the  ZAMS and  PG1159 stages,  respectively), 
and  the surface abundance  of the main
chemical  constituents during  the  PG1159 stage. The sequence  with
initial mass  of 2.7 $M_{\odot}$ is  the same presented  by Althaus et
al. (2005). 

\begin{table}
\centering
\caption{Initial and final stellar mass 
(in solar units), and the final 
surface chemical abundances by mass (PG1159 regime) for the evolutionary 
sequences considered in this work.}
\begin{tabular}{llccccc}
\hline
\hline
$M_{\rm ZAMS}$ & $M_{\rm PG}$ & $^4$He & $^{12}$C & $^{13}$C & $^{14}$N & $^{16}$O \\
\hline
 1    & 0.530 &  0.33  &  0.39 & 0.051  & 0.019  & 0.17  \\
 1    & 0.542 &  0.28  &  0.41 & 0.051  & 0.018  & 0.21  \\
 2.2  & 0.565 &  0.39  &  0.27 & 0.048  & 0.027  & 0.22  \\
 2.7  & 0.589 &  0.31  &  0.38 & 0.040  & 0.012  & 0.23  \\
 3.05 & 0.609 &  0.50  &  0.35 & 0.003  & 0.002  & 0.10  \\
 3.5  & 0.664 &  0.47  &  0.33 & 0.019  & 0.013  & 0.13  \\
 3.75 & 0.741 &  0.48  &  0.34 & 0.0007 & 0.0002 & 0.14  \\
\hline
\end{tabular}
\label{table2}
\end{table}  

Our PG1159 models are characterized by envelopes with uniform chemical 
compositions that extend from the surface downwards well below the 
driving region (i.e., the chemical composition at the driving region 
is the same 
than at the stellar surface). Thus, our models are not characterized by
chemical gradients between the driving region and the stellar 
surface. Note that our post-VLTP models  predict  a  range  in  the surface  
composition.  In particular, 
the final  surface abundance  of helium  spans the
range  0.28-0.50 by  mass\footnote{These abundances are not only determined by 
the stellar mass, but also by the deepness of the envelope 
convection and mass loss episodes after  the VLTP  (see
Miller Bertolami \&  Althaus 2006 for comments). Also, 
the number of thermal  pulses
considered in the AGB stage  and the efficiency of overshooting play a
role  in the  final  PG1159 surface  composition  (Herwig 2000).},  
which is  in  agreement with  the range  of
observed helium abundance in most PG1159s (see Werner \& Herwig 2006).
Our sequences  with helium abundances quite larger
than  the standard  ones observed  in PG1159  stars will  allow  us to
explore, at some extent, the role of helium  in the instability 
properties of pulsating
PG1159s. Mass loss episodes  after the VLTP have  not been considered  
in the PG1159 evolutionary sequences we employed here.

The pulsation stability analysis was performed 
with the help of a new finite-difference nonadiabatic  pulsation code 
which is based on the adiabatic version described  
in C\'orsico \& Althaus (2005, 2006). The nonadiabatic code solves the
full  sixth-order complex  system of linearized equations and boundary 
conditions as given by Unno et al. (1989). 
Our code provides the dimensionless  complex   
eigenvalue  ($\omega$)  and  eigenfunctions ($y_1, \cdots,  y_6$) 
as given  by Unno et al.   (1989).  Nonadiabatic
pulsation periods and normalized growth rates are evaluated as $\Pi= 2
\pi/ \Re(\sigma)$ and $\eta= -\Im(\sigma)/ \Re(\sigma)$, respectively.
Here, $\Re(\sigma)$ and $\Im(\sigma)$ are the real and the 
imaginary part, respectively, of the complex eigenfrecuency 
$\sigma=  (G   M_*/  R_*^3)^{1/2}\   \omega$. Our   code  
also  computes the  differential  work
function, $dW(r)/dr$, and the  running work integral, $W(r)$, as
in Lee \& Bradley (1993). In this work the 
``frozen-in  convection'' approximation was assumed because 
the flux carried by convection is usually negligible 
in PG1159 stars. Also, the $\epsilon$-mechanism for mode driving   
was neglected  in the computations because nuclear-burning shells in PG1159 
models usually  destabilize very  short  periods that are not observed 
in  GW Vir stars (Kawaler et  al. 1986; Kawaler 1988;
Gautschy 1997).  We employed about 3000 mesh-points to
describe our background stellar  models, most of them distributed
in the  envelope region where  all the pulsation driving  and damping
occur.   We employed the ``Ledoux modified''  treatment  to compute the 
Brunt-V\"ais\"al\"a   frequency  ($N$) (Tassoul  et  al.   1990).

\begin{figure}
\centering
\includegraphics[clip,width=250pt]{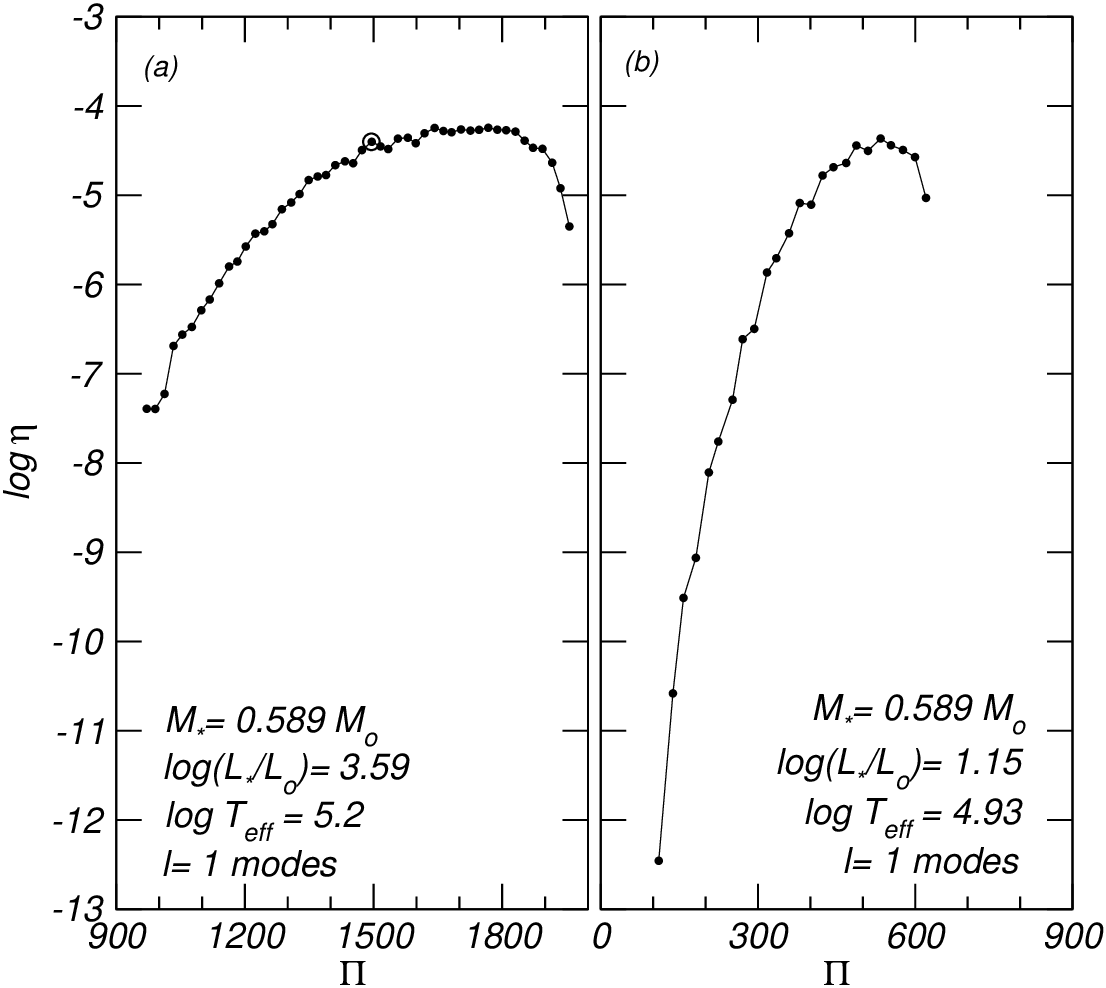}
\caption{The normalized growth rate in terms of period (in seconds) 
for overstable $g$-modes corresponding to 
two $0.589 M_{\odot}$ PG1159 models. 
The location of these models on the HR diagram is shown in 
Fig. \ref{track}. Circumscribed dot in panel (a) 
correspond to a mode with  $k= 70$.}
\label{growth}
\end{figure}

\section{Stability calculations}
\label{s-calculations}

We analyze the stability  properties of about
2400 stellar  models covering a  wide range of  effective temperatures
($5.4 \gtrsim \log  (T_{\rm eff}) \gtrsim 4.8$) and  a  range of
stellar  masses ($0.530  \lesssim M_*/M_{\odot}\lesssim  0.741$).  For
each  model we  have restricted  our study  to unstable  $\ell=  1, 2$
$g$-modes with periods in the range $ 50\ {\rm s} \lesssim \Pi \lesssim 
7000$ s, thus comfortably embracing 
the full period spectrum observed in GW Vir stars.  
In line with  other stability studies of  GW Vir
stars, all unstable  $g$-modes in our PG1159 models  are driven by the
$\kappa$-mechanism  associated with  the opacity  bump due  to partial
ionization  of  K-shell  electrons   of  C$_{\rm  V-VI}$  and  O$_{\rm
VII-VIII}$  centered at  $\log T  \approx 6.2$  (Quirion et  al. 2004;
GAS05).
 
We start by discussing the stability properties of two template $0.589
M_{\odot}$ PG1159 models. These properties are common
to all  PG1159 models of  our complete set of  evolutionary sequences.
The  normalized growth  rate ($\eta$)  in terms  of  pulsation periods
($\Pi$)  for  overstable $\ell=  1$  modes  corresponding  to the  two
selected  models  are  shown  in  Fig.  \ref{growth}.   Model  (a)  is
representative of the high-luminosity, low-gravity pre-white dwarf
regime, and  model (b) is typical  of the low-luminosity, high-gravity 
phase,  when the object  has already entered their white
dwarf  cooling  track  (see  Fig. \ref{track}).   Note  that  modes
excited  in model  (a) have  pulsation  periods in the range 
$  1000 \lesssim  \Pi \lesssim 2000$ s, substantially longer than 
those excited in model (b) ($ 100 \lesssim  \Pi \lesssim 600$ s).  
For  each model,  $\eta$ reaches  a maximum
value in the  vicinity of the long-period boundary  of the instability
domain. In  other words,  within a given  band of unstable  modes, the
excitation  is  markedly  stronger  for modes  characterized  by  long
periods.  This  effect is particularly  notorious in model  (b), being
the value of the growth rate  for the shortest periods more than seven
order  of magnitude  smaller than  for the  modes with  longer periods
\footnote{We note,  however, that for modes with  the shortest periods
the value  of $\eta$ is so  small ($\lesssim 10^{-9}$)  that they are
only  marginally  unstable.}.

\begin{figure}
\centering
\includegraphics[clip,width=250pt]{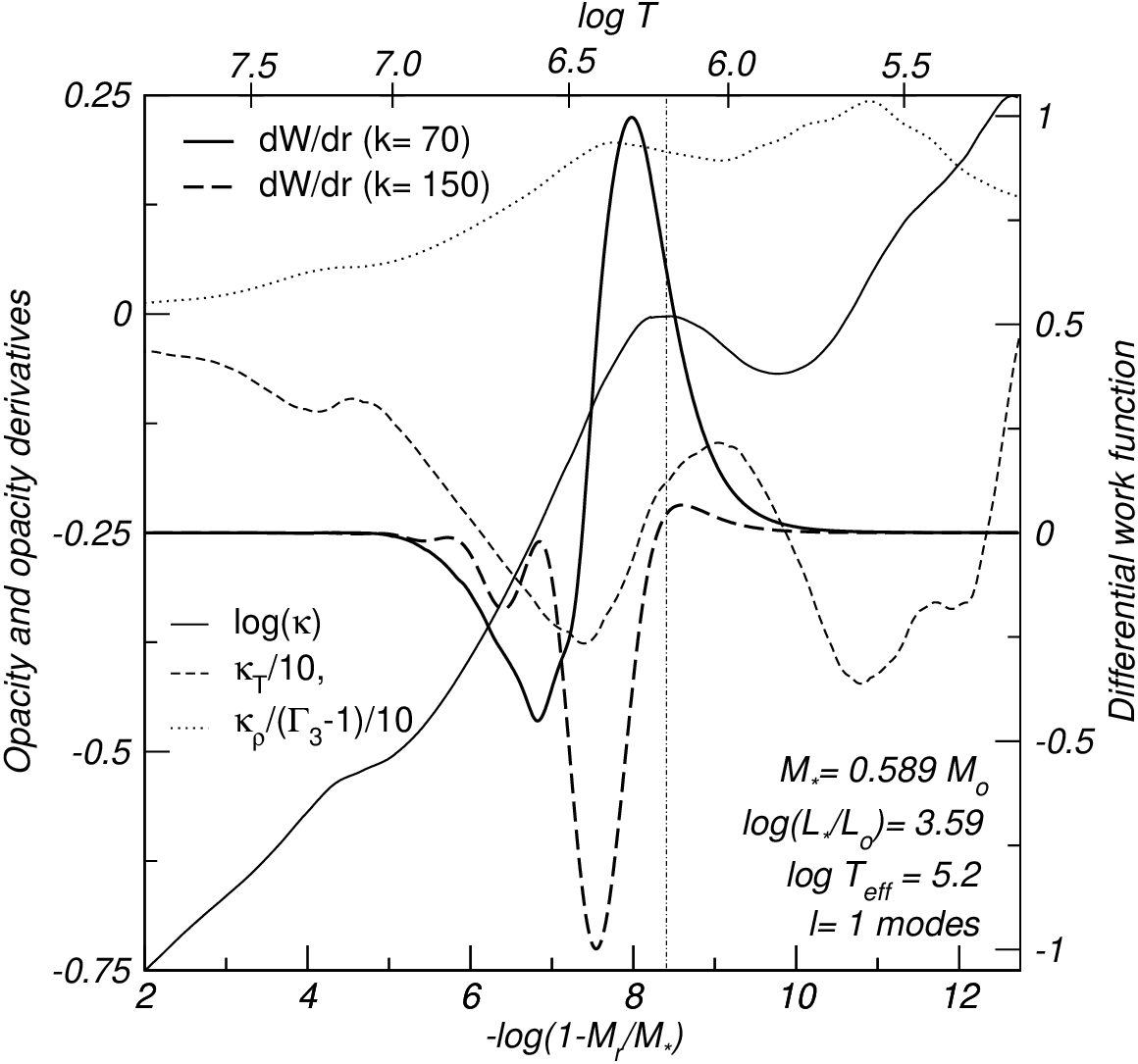}
\caption{The opacity and opacity derivatives (left scale), 
and the differential work function (right scale) 
for two selected $\ell= 1$ modes in terms of the mass coordinate 
(lower scale) and the logarithm of the temperature (upper scale), 
corresponding to the PG1159 model (a) analyzed in Fig. \ref{growth}.
$dW/dr$ for the mode with $k=70$ ($k= 150$) 
is normalized to an extremum value of $+1$ ($-1$). The vertical  
dot-dashed line shows the location of the maximum in the 
C/O opacity bump.}
\label{opa1}
\end{figure}

Fig. \ref{opa1} shows details  of the driving/damping process in model
(a) for two selected dipole modes.  We restrict  the figure to the 
envelope region of the model, where the main driving and damping occurs.  
Thick continuous curve  corresponds to $dW/dr$ for an unstable mode with
$k= 70, \Pi=  1496$ s and $\eta= 3.9 \times  10^{-5}$ 
(marked as a circumscribed dot in panel (a) of Fig. \ref{growth}), 
while the thick dashed one depicts the situation for a 
stable mode  with $k= 150, \Pi= 3216$ s, $\eta= -1.23 \times 10^{-3}$.  
Also plotted is the run of the Rosseland  opacity, $\kappa$,   and the  
run of  its logarithmic  derivatives,  $\kappa_{\rm  T}$  and
$\kappa_{\rho}/(\Gamma_3-1)$.  As can be seen, the
region  that destabilizes the  $k= 70$  mode (where  $dW/dr >  0$) is
clearly associated with  the bump in the opacity  centered at $-\log q
\approx 8.4$  [$q \equiv  (1-M_r/M_*)$], although the  maximum driving
for this  mode comes  from a slightly  more internal region  ($-\log q
\approx  8$).  Note  also  that  in the  driving  region the  quantity
$\kappa_{\rm T} +
\kappa_{\rho}/(\Gamma_3-1)$ is increasing  outward, in agreement 
with the well known necessary condition for mode excitation 
(Unno et al. 1989).
Since the contributions to driving
at $-\log q$ from 7.5 to  10 largely overcome the damping effects at $
6 \lesssim  -\log q \lesssim 7.5$,  the mode with $k=  70$ is globally
excited. At variance, the strong damping experienced by the mode with $k=
150$ (denoted by negative values of $dW/dr$), makes this mode globally
stable. The situation at the low-luminosity, high-gravity 
phase as in model (b) of Fig. \ref{growth} is
very similar,  the  only important  difference  being that  the
driving/damping  regions  are located  at considerably
more external  layers. This is  due to  an outward
migration of the opacity profile, induced by the evolution of the star.

\subsection{The theoretical GW Vir instability strip}
\label{location}

\begin{figure}
\centering
\includegraphics[clip,width=250pt]{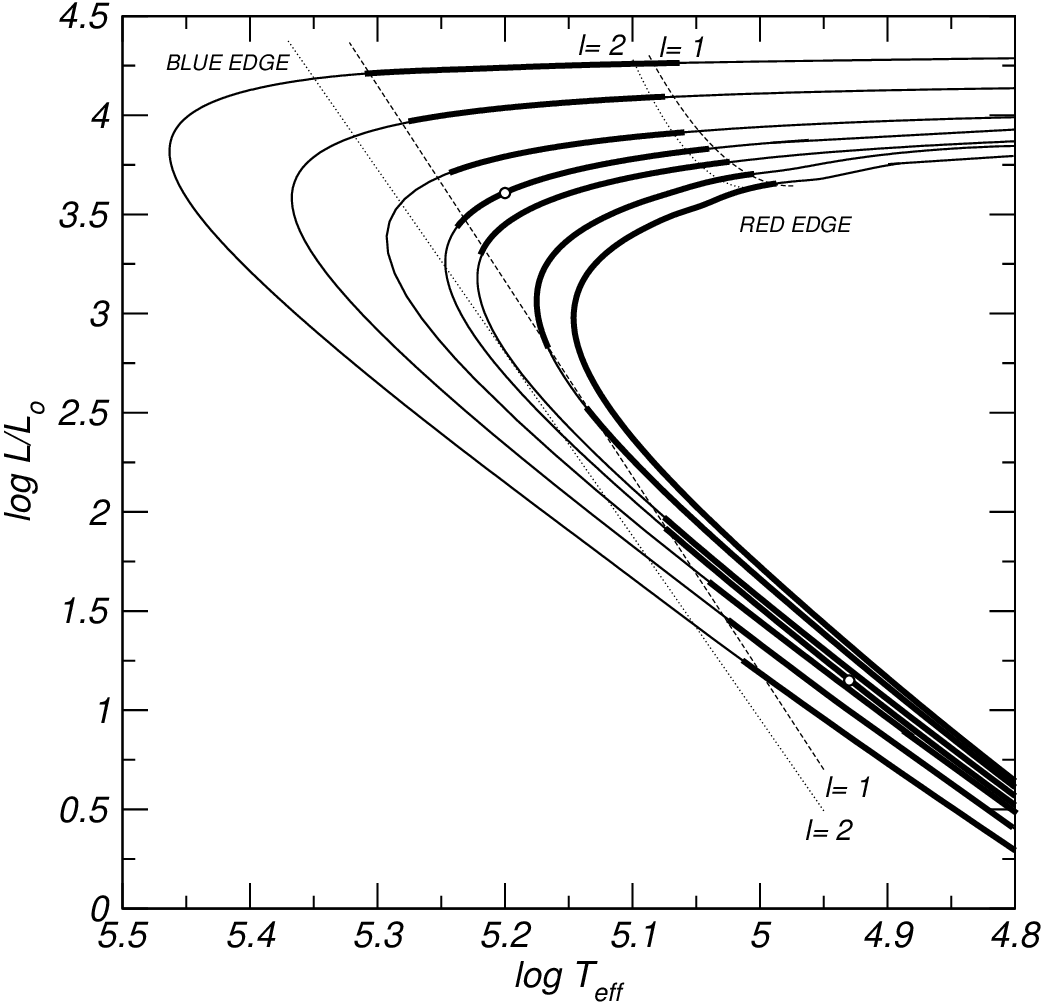}
\caption{The HR diagram for our complete set of PG1159 model sequences
of $0.530, 0.542,  0.565, 0.589, 0.609, 0.664$ 
and $0.741 M_{\odot}$ (from right to left).
The portions of the tracks harbouring stellar models 
with overstable $\ell= 1$ $g$-modes are 
emphasized with thick solid lines. Thin dashed ($\ell= 1$) and 
dotted ($\ell= 2$) lines are  
parameterizations of the blue and red edges of instability. 
The small white circles on the 
$0.589 M_{\odot}$ track show the loci of the two template models 
analyzed in Fig. \ref{growth}.}
\label{track}
\end{figure}

Here, we examine the location of the unstable  domains on the HR
diagram. In Fig.   \ref{track} we  show  
the evolutionary tracks for  our complete set  
of  PG1159 model sequences, where the thick portions of
the curves  correspond to models with dipole unstable modes.
A well-definite instability domain, bounded by  a red (cool) 
edge at high luminosities, and by a  blue (hot) edge both at high 
and  low luminosities, is apparent in the plot.  
The blue  and  red  edges for dipole and quadrupole modes 
for each sequence are connected by thin 
curves as given by standard nonlinear least-squares  algorithms.  
The instability domains for $\ell= 1$ and $\ell= 2$ look 
very similar, although  the  edges for $\ell= 2$ are  
slightly shifted  to  higher effective  temperatures,  
and the  region  of  instability is  somewhat wider than for $\ell= 1$. 
Fig.   \ref{track} should be compared with Fig.  5 of GAS05. 
The global agreement between our results and  the predictions 
of GAS05 is  excellent, in particular for the sequence of 
$0.589  M_{\odot}$ --- the only sequence in common
between those authors  and our work.  

At variance  with GAS05, in this  work we have  employed PG1159 models
with different  masses derived from the complete  evolution of the
progenitor stars. This has enabled us to extend the
pulsational  stability  analysis to  lower effective temperatures in the 
high-luminosity, low-gravity region. As  a
result, we  have been able to found, for the first time,  
a reliable high-luminosity, low-gravity red edge  of the GW
Vir instability strip. Clearly, the red edge is markedly sensitive to
the stellar mass,  being more hotter for the  more massive models. 

Our blue edge (both for dipole and quadrupole modes) 
cannot  be exactly  represented by a straight line. It is clearly 
seen for dipole modes in Fig. \ref{track}.  
The  departures from  a simple  linear  relation have  
their origin in the different surface  chemical compositions 
with which our models of different stellar masses 
reach the domain of the PG1159 after emerging 
from the born-again episode.  In fact,  the
unstable portions  of the evolutionary tracks  corresponding to models
characterized by a surface  helium abundance of $\approx 0.28-0.39$ by
mass (that is, the sequences  with masses of $0.542, 0.565$ and $0.589
M_{\odot}$) extend  slightly beyond the linear parameterization of  the blue
edge, as compared with the case  of the more massive models which have
larger  helium abundances in  the envelope  ($0.50, 0.47$ and $0.48$ for
stellar masses of $0.609, 0.664$ and $ 0.741 M_{\odot}$, 
respectively)\footnote{Note that 
the helium abundance range predicted by our models ($0.3-0.5$)
is coincident with the observed one  for most of PG1159 stars, as reported 
by Werner \& Herwig (2006).}. 
In fact, increasing the helium abundance at the 
driving region the efficiency of  pulsational driving  is 
reduced  (see GAS05). This is in line also with the finding of 
Quirion et al. (2004) that decreasing  the helium mass  fraction at  
the driving regions,  the blue  edge of  the instability  domain 
shifts  to higher effective  temperatures. 

\begin{figure}
\centering
\includegraphics[clip,width=250pt]{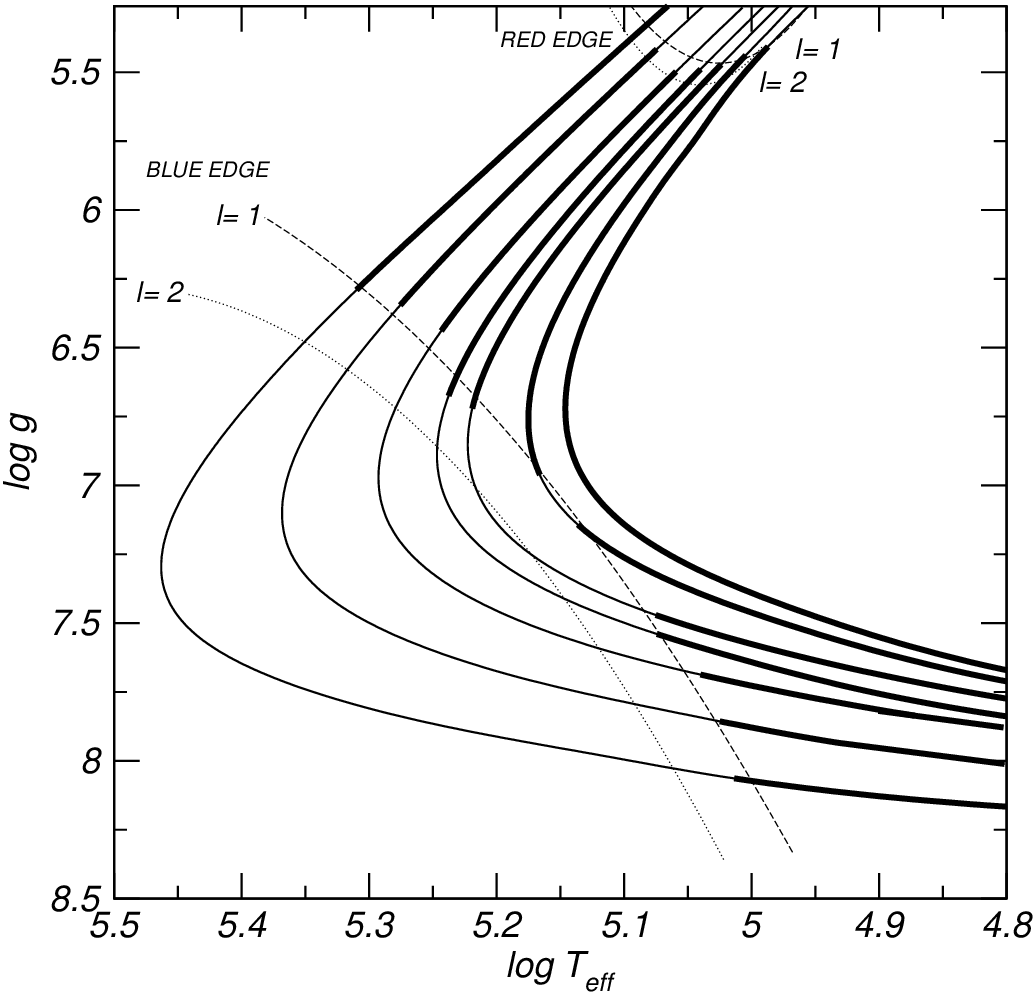}
\caption{Same as Fig. \ref{track}, but for the  
$\log (T_{\rm eff}) - \log g$ plane.} 
\label{grave}
\end{figure}

Since  spectroscopic  calibration  of  PG1159  stars  gives  effective
temperatures  and surface  gravities,  it  is useful  to  see how  the
evolutionary tracks  and the instability  domains look on the  $\log (T_{\rm
eff}) -  \log g$ plane.  Fig. \ref{grave} plots the 
complete  set of PG1159 sequences on that  plane, 
in which we emphasize with thick solid lines the stages
with overstable $\ell= 1$ $g$-modes.
Again, we  characterize the loci of  the blue and red  edges for dipole 
and quadrupole modes with thin curves  as   parameterized  by     
standard  nonlinear  least-squares
procedures.  We postpone to \S \ref{theory-observations} 
a  complete discussion  of the general agreement between our pulsation 
models and the observed GW Vir stars.

In  the current  computations,  the dipole  unstable  domain for  each
evolutionary  sequence  is  separated  into  two regions,  one  of  them
corresponding to the high-luminosity phase (low gravity) and the other
corresponding    to    low    luminosities   (high    gravity)    (see
Figs. \ref{track}  and \ref{grave}).  The only  exception is the
sequence  with $0.530  M_{\odot}$  which exhibits  an unique  unstable
domain that extends uninterruptedly  from the high-luminosity phase to
the  low-luminosity regime.   This  is at  variance  with the  results
reported by  GAS05, who  found that for  all of their  sequences (including
that  with $0.530  M_{\odot}$) the  tip  of the  evolutionary ``knee''  is
pulsationally stable in the case  of dipole modes. This discrepancy is
connected with the different loci of  the tracks on the HR diagram and
on the  $\log (T_{\rm eff}) -  \log g$ plane. In  fact, the evolutionary
sequences  considered in  this work  reach lower  $\log g$  and
luminosity values for a  fixed  $\log (T_{\rm  eff})$  than in  the ones  of
GAS05. This reflect the fact that in the present work we have considered 
PG1159 evolutionary sequences derived from the full evolutionary 
computations of  the progenitor stars.

\begin{figure}
\centering
\includegraphics[clip,width=250pt]{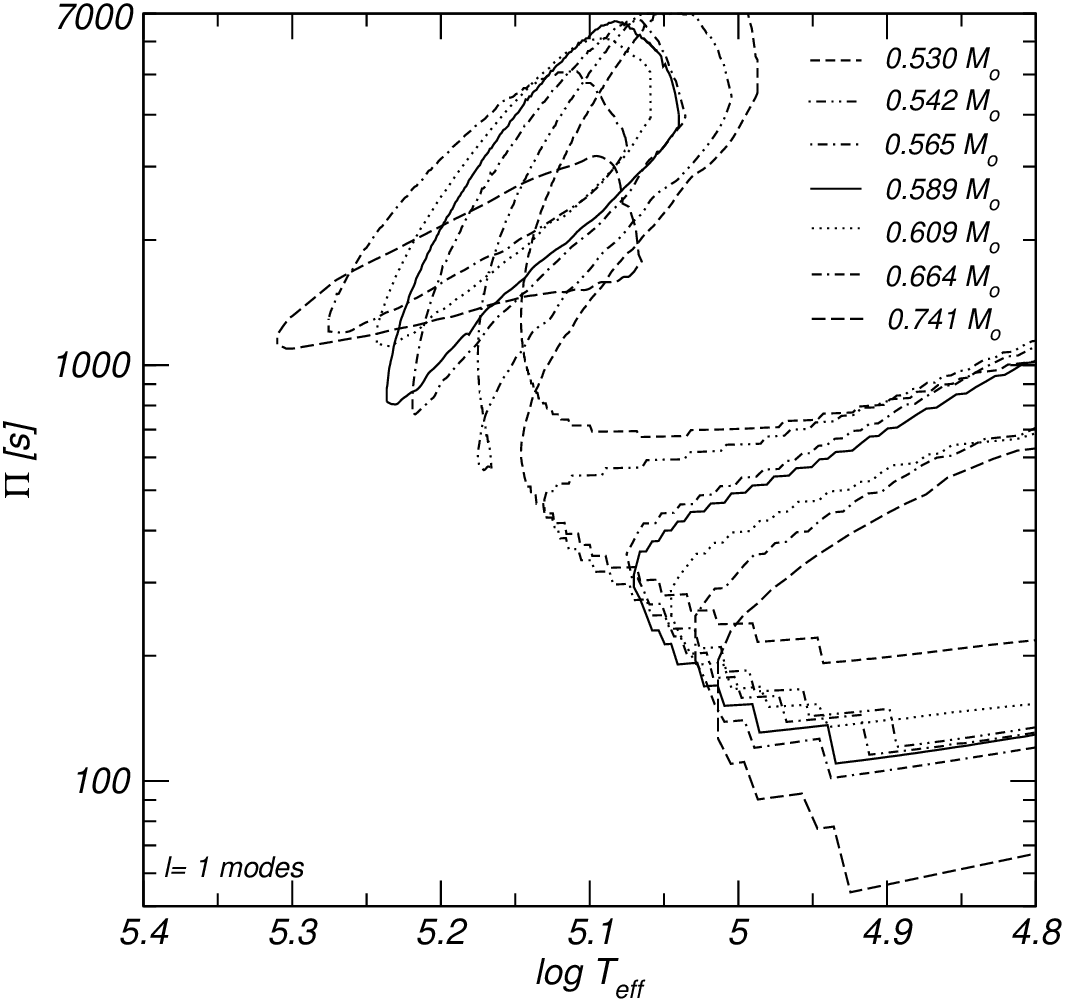}
\caption{The instability domains on the $\log (T_{\rm eff})-\Pi$
plane for $g$-modes with $\ell= 1$ corresponding to our complete set 
of PG1159 model sequences. Different line styles depict the 
situation for different stellar masses.}
\label{cont-l1}
\end{figure}

We explore now the ranges of  periods of unstable modes.  We
begin  by  examining  Fig.   \ref{cont-l1}  which  shows  the
instability boundaries of dipole  modes on the $\log(T_{\rm eff})-\Pi$
plane  for the complete  set of  evolutionary sequences.  
Note  that generally the periods of unstable  modes for each sequence 
are clearly grouped into two separated regions,  one of them characterized 
by long periods and high-radial  overtones, corresponding to evolutionary 
stages before the knee at  high  luminosities,  and the  other one
characterized by  short periods (low $k$ values)  corresponding to the
hot portion  of the white  dwarf cooling track (low  luminosities). As
mentioned  before, the  sequence  with $0.530  M_{\odot}$ exhibits  an
unique   instability  domain.   Thus,  this
sequence    shows   instability    even    along  the whole  stages
around the knee. The splitting of the instability domains into two 
separated regions can be understood in terms of the magnitude of 
the thermal timescale 
at the driving region. We refer the reader to the paper by GAS05 for an 
illuminating demonstration of this. 

The  high-luminosity  domain of  instability  exhibits  a strong  mass
dependence, being the longest  unstable periods  shorter  for the  
more  massive models.  In addition,  for the  less massive  models the  
longest pulsation period is reached at lower effective temperatures as 
compared with the
situation  for  the  more  massive   models.  In  all  the  cases  the
long-period  limit is  attained  shortly after  the  beginning of  the
instability domain.  The shortest  unstable period for each sequence is
also markedly  sensitive to the stellar mass.   Generally, the minimum
period  is smaller  for  the less  massive  models. Thus,  
the period-width  of the  instability  domain is
larger   for  the   less  massive   models.  As   can  be   seen  from
Fig.   \ref{cont-l1},  the  instability   island  on   the  $\log(T_{\rm
eff})-\Pi$ plane is almost vertical for the model with $0.530 M_{\odot}$, but
its  slope gradually decreases as  we go to  sequences with higher
masses.

The  low-luminosity  domain,  on  the  other hand,  shows  a  moderate
dependence  on the stellar  mass. The  maximum overstable  period is
always  larger for the  less massive  models.  The  minimum overstable
period, however,  does not show a  clear trend with the  value of the
stellar  mass. We  note  that  the shortest  unstable  periods are  of
$\approx  55$   s  and  correspond   to  the  sequence   with  $0.741
M_{\odot}$. This  short-period limit is substantially  lower than that
reported  by  GAS05,  of about  190  s  for  their sequence  of  $0.64
M_{\odot}$.  It is  important to  note  that in  our computations  the
shortest overstable periods have very  small growth rates, of the order
of $10^{-9}- 10^{-  13}$, and thus we could  consider that these modes
are  stable  [see  panel  (b)  of  Fig.  \ref{growth}].   If  so,  our
short-period limit agrees with the predictions of GAS05. 
The morphology  of the instability  domains for quadrupole  modes looks
very similar  to the case of  dipole modes, being the  novel feature a
markedly  shortening  of the  overstable  periods,  in agreement  with
GAS05.    Indeed,  the  long-period   limits  of   the  high-luminosity
instability domains for quadrupole modes are shortened in about 3000 s as
compared with  the case  of dipole ones.  A less severe  decreasing of
overstable quadrupole periods  is also present  
in the  low-luminosity domain.  As a 
result, both regions  of instability are      closer  between them, to
such a  degree that the high- and  low-luminosity instablility domains
of the  $0.530, 0.542$ and  $0.565 M_{\odot}$ sequences merge  into an
unique, uninterrupted  region\footnote{Note that in Figs. \ref{track} 
and \ref{grave} the blue edge for $\ell= 2$ crosses the evolutionary track 
of $0.565 M_{\odot}$; we stress that it is simply a {\it parameterization} 
of the blue edge and not the {\it exact} theoretical  blue edge
emerging from our stability analysis.}. 
The  lowest $\ell= 2$ short-period limit  for the
low-luminosity  domain, which  corresponds to  the sequence  with 
$M_*= 0.741 M_{\odot}$, is  of about  50 s. 

In all of the computations presented in this  paper we 
neglected  chemical diffusion, and thus  the stability  calculations were
performed on PG1159 models with a constant chemical composition at the
driving region. In particluar,  no helium  enrichment  at the driving region 
was allowed,  and consequently all of our  sequences shown 
pulsational instability well beyond the 
empirical red  edge of the GW  Vir stars at low luminosities 
($T_{\rm eff}  \sim 75\,000$ K; Dreizler \& Heber 1998), even 
down to the domain of the variable DB white dwarfs.

\section{Theory vs observations}
\label{theory-observations}

\begin{table*}
\centering
\caption{Stellar parameters and pulsation properties 
of all known pulsating PG1159 stars. The approximate surface abundances (in 
\% by mass) have been derived from Table 2 of Werner \& Herwig (2006)
by assuming a composition made of $^{4}$He, $^{12}$C and $^{16}$O,
except for HS 2324+3944 (data taken from Table 1 of Werner \& Herwig 2006) 
and for Abell 43 (data taken from Miksa et al. 2002).}
\begin{tabular}{rccccccrcc}
\hline
\hline
 Star & $\log (T_{\rm eff})^{\rm a}$ & $\log g^{\rm a}$ & 
$M_*^{\rm b}$ &  H & He & C & O & Remark & Observed period range ($\ell= 1$)\\
& [K] & [cgs] & [$M_{\odot}$]&  &  &  &  & & [s] \\

\hline
 Abell 43      & 5.041 & 5.7 & 0.53 & 35 & 42 & 23  &  -     & PNNV, hybrid  & $2600-3035\ ^{\rm c}$  \\
 HS 2324+3944  & 5.114 & 6.2 & 0.53 & 17 & 35 & 42 &  6 & naked, hybrid &  $2005-2569\ ^{\rm d}$ \\
 Longmore 4    & 5.079 & 5.5 & 0.63 & -  & 48 & 43 &  9 & PNNV          & $831-2325\ ^{\rm e}$   \\
 NGC 246       & 5.176 & 5.7 & 0.75 & -  & 63 & 31 &  6 & PNNV          & $1460-1840\ ^{\rm f}$  \\
 K 1-16        & 5.146 & 6.4 & 0.54 & -  & 33 & 50 & 17 & PNNV          & $1500-1700\ ^{\rm g}$  \\
 RX J2117+3412 & 5.230 & 6.0 & 0.72 & -  & 39 & 55 &  6 & PNNV          & $694-1530\ ^{\rm h}$   \\
 HE 1429-1209  & 5.204 & 6.0 & 0.66 & -  & 39 & 55 &  6 & naked         & $919\ ^{\rm i}$        \\
 PG 1159-035   & 5.146 & 7.0 & 0.54 & -  & 33 & 50 & 17 & naked         & $430-840\ ^{\rm j}$    \\
 PG 2131+066   & 4.978 & 7.5 & 0.55 & -  & 44 & 39 & 17 & naked         & $339-598\ ^{\rm k}$    \\
 PG 1707+427   & 4.929 & 7.5 & 0.53 & -  & 44 & 39 & 17 & naked         & $335-909\ ^{\rm l}$    \\
 PG 0122+200   & 4.903 & 7.5 & 0.53 & -  & 44 & 39 & 17 & naked         & $336-612\ ^{\rm m}$    \\
\hline
\end{tabular}

{\footnotesize References: $^{\rm a}$ Werner \& Herwig (2006); 
$^{\rm b}$ Miller Bertolami \& Althaus (2006);
$^{\rm c}$ Vauclair et al. (2005); $^{\rm d}$ Silvotti et al. (1999); $^{\rm e}$ Bond \& Meakes (1990); 
$^{\rm f}$ Ciardullo \& Bond (1996); $^{\rm g}$ Grauer et al. (1987); $^{\rm h}$ Vauclair et al. (2002); 
$^{\rm i}$ Nagel \& Werner (2004); $^{\rm j}$ Winget et al. (1991); $^{\rm k}$ Kawaler et al. (1995);
$^{\rm l}$ Kawaler et al. (2004); $^{\rm m}$ O'Brien et al. (1998).} 
\label{gwvirginis}
\end{table*}  

In  this  section we  compare  our  theoretical  predictions with  the
observed properties  of  GW  Vir stars.   Currently,  11 pulsating
PG1159 stars  are presently known.  In Table  \ref{gwvirginis} we show
the main  spectroscopic and pulsation data available.  Note that there
are  five GW  Vir stars  (termed PNNV)  that are  still embedded  in a
planetary nebula.  The remainder objects lack a surrounding nebula and
are commonly  called ``naked''  GW Vir stars.  Finally, there  are two
objects with measurable amounts of hydrogen in their spectra; they are
termed pulsating ``hybrid-PG1159''. Note  that HE 1429-1209 is a naked
GW Vir star but its effective  temperature and gravity place it at the
region of the HR diagram usually populated by PNNVs.
 
\begin{figure}
\centering
\includegraphics[clip,width=250pt]{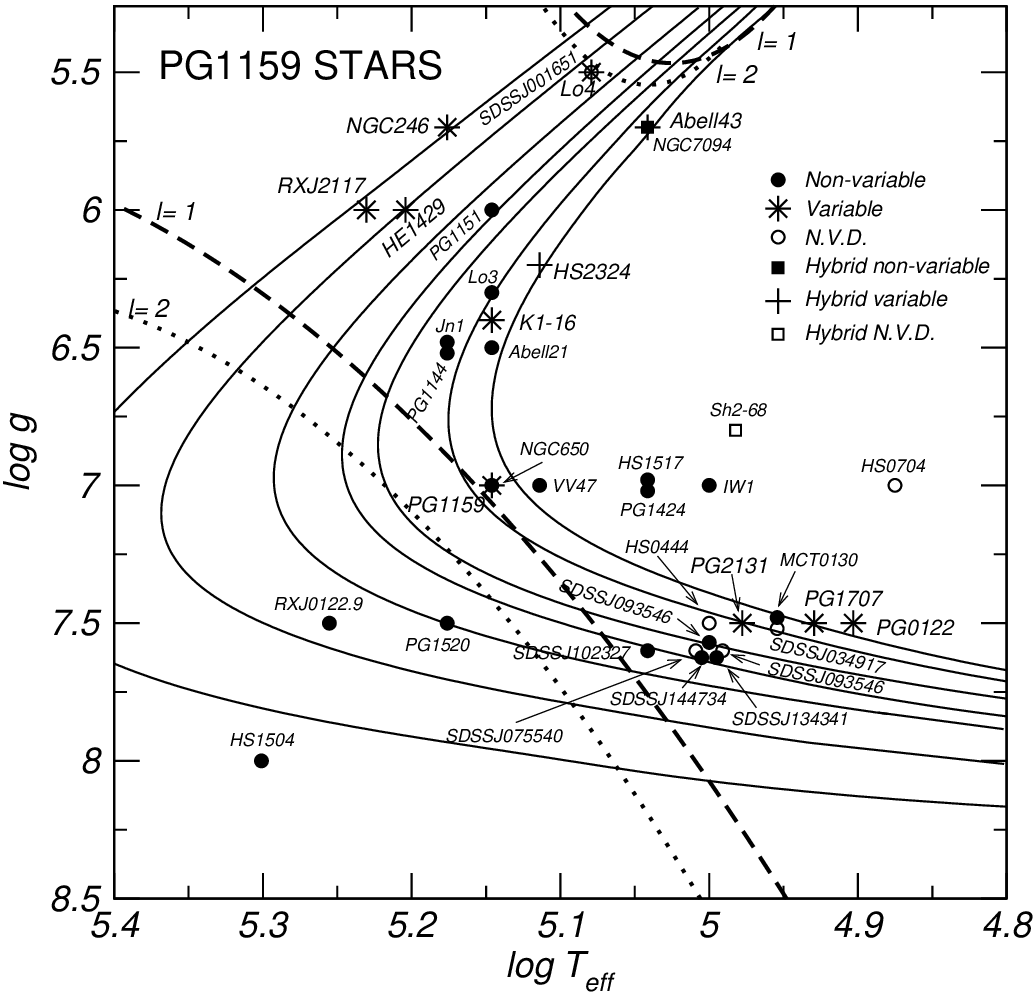}
\caption{The distribution of the spectroscopically calibrated variable  
and non-variable PG1159 stars 
in the  $\log (T_{\rm eff}) - \log g$ 
plane. PG1159 stars with no variability data (N.V.D.) are depicted with 
hollow circles. Solid curves show the evolutionary tracks for 
different stellar masses: $0.530, 0.542,  0.565, 0.589, 0.609, 0.664$ 
and $0.741 M_{\odot}$. 
Parameterizations of the theoretical dipole 
(dashed curves) and quadrupole 
(dotted curves) red and blue edges of the 
instability domain are also displayed.} 
\label{grave-obs}
\end{figure}

In  Fig.   \ref{grave-obs}  we  plot  the location  of  pulsating  and
nonpulsating PG1159 stars, as well  as PG1159 stars that have not been
observed for variability,  on the $\log (T_{\rm eff})  - \log g$ plane
(data  taken from  Werner \&  Herwig 2006).  The plot  also  shows the
evolutionary  tracks  for our  complete  set  of  sequences of  PG1159
models. The  blue and  red edges for  dipole and quadrupole  modes are
also shown.   Regarding pulsating PG1159 stars,  the agreement between
observations and  model predictions is excellent.  In  fact, {\it all}
of the GW Vir variables lay inside our predicted instability  domains of dipole
and quadrupole  modes.  We also  see that, however, there  are several
non-variables  occupying   the  unstable  region.   The  existence  of
non-variable  stars within  the instability  domain could  in  part be
understood  in terms of  a variation  in surface  chemical composition
(and  thus  in  the  driving  region)  from  star  to  star.  
For instance, Quirion et al. (2004) found that the helium enrichment at the
driving region is  the cause for the existence of the nonpulsator 
MCT 0130-1937 (with a helium abundance of about 75 \%) 
within the  instability  strip.  
Note,  however,  that  PG  1151-029,
Longmore 3, Abell 21 and VV47 (not included in the analysis of Quirion
et al. 2004) are found  to have standard helium abundances (see Table
2  of Werner \&  Herwig 2006)  and however  are non-variables.  On the
other hand, it is remarkable the existence of the pulsating star 
NGC 246 with a helium abundance ($X_{\rm
He}  \approx  0.63$)  unusually   large  among  pulsators  (see  Table
\ref{gwvirginis}). These controversial cases remain to be explained.

Regarding PG 1159-035 (GW Vir)  --- the prototype of both the variable
and the spectroscopic classes  --- our analysis naturally predict that
this  star  {\it does  pulsate}  for a  stellar  mass  of $\sim  0.536
M_{\odot}$ (see Fig.  \ref{grave-obs}).  It is a markedly higher value
than the  mass required by GAS05  to excite pulsations  in PG 1159-035
($M_* \approx 0.50  - 0.51 M_{\odot}$), and, at  the same time, closer
to  the  long-recognized   ``seismic''  mass  of  $\approx  0.59
M_{\odot}$ of Kawaler \& Bradley (1994).  So, the employment of PG1159
models derived  from the complete evolution of  their progenitor stars
appears to  be a key factor  to alleviate the  discrepancy between the
seismic and  the spectroscopic mass  of PG 1159-035.  

\begin{figure}
\centering
\includegraphics[clip,width=250pt]{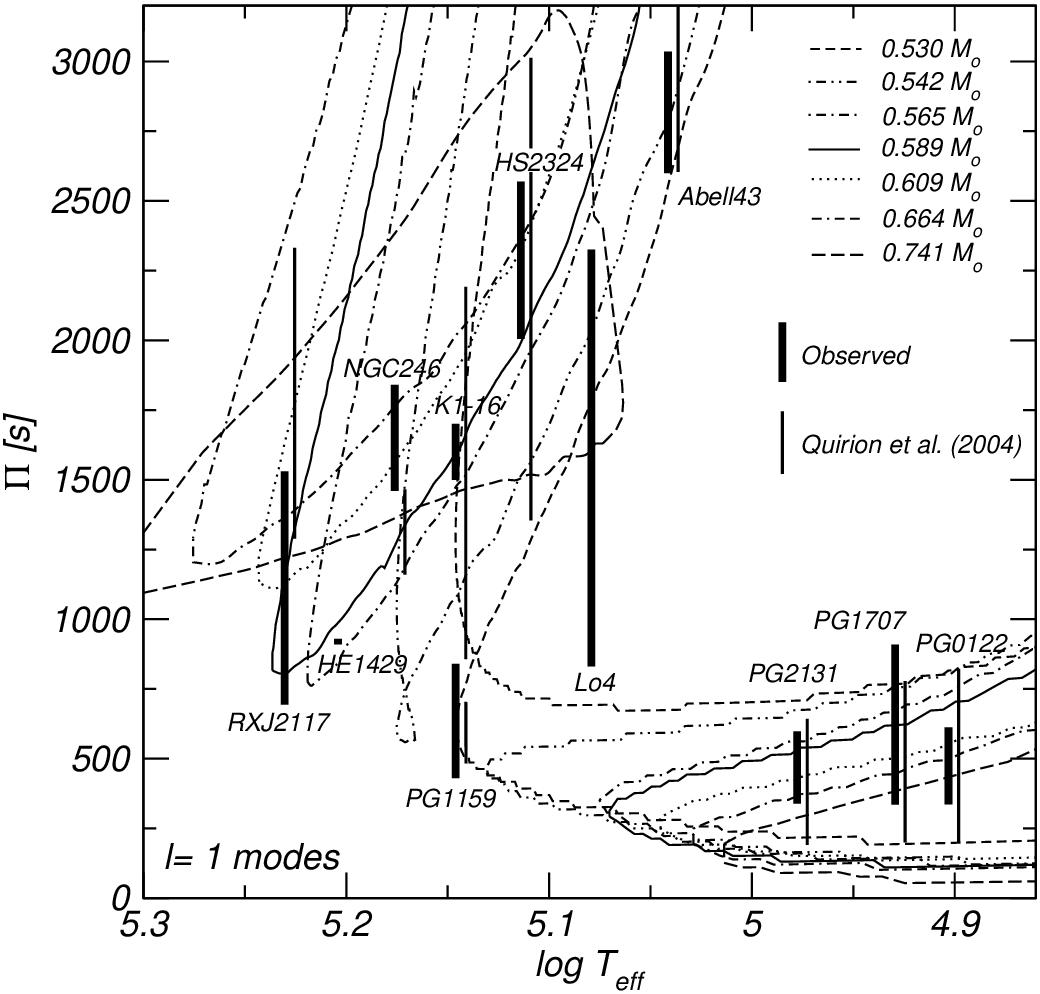}
\caption{The period ranges of known 
variable PG1159 stars on the $\log(T_{\rm eff})-\Pi$
plane, depicted with thick vertical bars (the thickness of the bars
is arbitrary). For comparison, we include 
the predicted ranges of unstable periods according to Quirion et al. (2004),
shown with thinner vertical bars (arbitrarily shifted to right 
by 0.05 dex for clarity). The  
theoretical instability domains for $g$-modes with $\ell= 1$ 
for our complete set of PG1159 sequences are depicted 
with different line styles.} 
\label{cont-l1-obs}
\end{figure}

In what follows  we shall focus on the observed  period ranges for the
known GW Vir stars.  In Fig.  \ref{cont-l1-obs} we show the  $\log(T_{\rm
eff}) -\log \Pi$ diagram, in  which the effective temperatures and the
period ranges are taken from Table
\ref{gwvirginis}. For comparison, we have included the theoretical 
instability boundaries for $\ell= 1$.   Also plotted are the ranges of
unstable periods  as predicted by  Quirion et al. (2004).   We stress
that we  are only interested  in a qualitative comparison  between the
observed and the theoretical ranges  of unstable periods.  Thus, we do
not  attempt here  a detailed  asteroseismic period  fitting for
each   individual    GW   Vir   star.     As   can   be    seen   from
Fig. \ref{cont-l1-obs}, the  agreement between theory and observations
is reasonably good for all  cases, and for several stars the agreement
turns out to be  excellent. Note, for  instance, that for seven  stars (Abell
43, HS  2324+3944, K1-16,  NGC 246, HE  1429-1209, PG 2131+066  and PG
0122+200) the  observed ranges of periods are  completely contained in
the  theoretical instability  domains  corresponding to  at least  one
stellar mass.  For  the case of PG 1159-035,  the observed periods are
well compatible  with the instability  domain of the  $0.53 M_{\odot}$
models, although  the shorter detected periods ($\lesssim  530$ s) lay
below the theoretical short-period limit.  The opposite is true for PG
1707+427, whose  longer observed periods  ($\gtrsim 760$ s)  lay above
the  highest long-period  limit  which corresponds,  at the  effective
temperature of this  star, to the $0.530 M_{\odot}$  sequence.  In the
case  of  RXJ  2117+3412,  the  range of  observed  periods  partially
overlaps with four instability regions, corresponding to the sequences
of $0.5890, 0.609,  0.664$ and $ 0.741 M_{\odot}$,  but no instability
domain accounts  for the observed  periods shorter than  $\approx 800$
s. Finally, in the case of  Longmore 4 the band of observed periods is
partially covered by the instability domains of the $0.530, 0.542$ and
$0.741  M_{\odot}$  sequences. Note  that  the  agreement between  the
observed and predicted periods of unstable modes is comparable to that
reported by Quirion et al. (2004).

The predicted  instability domains resulting from  our analysis nearly
reproduce  the  observational  trend  that the  periods  exhibited  by
pulsating PG1159 stars decrease with decreasing luminosity (increasing
surface  gravity)(see  O'Brien  2000   for  a  discussion  of  such  a
trend). This is well  documented by Figs. \ref{cont-l1} and 
\ref{cont-l1-obs}.  Note that Longmore 4,  being a high-luminosity
PNNV that shows short periods, is an exception to this trend. Longmore
4 is particularly interesting because it showed a surprising behaviour
in its spectral  type which suddenly changed from  PG1159 to [WCE] and
back again to  PG1159.  According to Werner et  al. (1992), this could
be a result  of a transient but significant increase  in the mass loss
rate. Quite  interestingly, according to our  computations, Longmore 4
is located  very close to the  red edge of the  instability strip (see
Fig.  \ref{grave-obs});  so, this  star  could  have  just entered  the
instability phase.

In  addition to  the  issue of    pulsation instability,
valuable  information  about  pulsating   PG1159  stars  can  be  also
extracted  from the  period spacing  between consecutive  overtones of
fixed  $\ell$ value.   The  period spacing  depends  primarily on  the
stellar  mass and  is  only  weakly dependent  on  the luminosity  and
surface  compositional   stratification  (Kawaler  \&   Bradley  1994;
C\'orsico  \& Althaus  2006). Thus, this 
quantity allows a determination of  $M_*$ to a very high  accuracy.  
Here, we consider five GW Vir stars  for which detailed 
asteroseismic studies have been carried out: 
PG1159-035 ($\overline{\Delta \Pi}= 21.5$ s; Kawaler
\& Bradley  1994), PG 2131+066 ($\overline{\Delta \Pi}=  21.6$ s; Reed
et al. 2000), PG 1707+427 ($\overline{\Delta \Pi}= 23.0$ s; Kawaler et
al. 2004),  PG 0122+200 ($\overline{\Delta  \Pi}= 21.1$ s;  O'Brien et
al.  2000),  and  RX  J2117+3412  ($\overline{\Delta  \Pi}=  21.5$  s;
Vauclair et al. 2002).  We compare the observed mean period spacing of
each  star   with  the  average   of  the  computed   period  spacing,
$\overline{\Delta    \Pi_k}$   ($\Delta    \Pi_k=   \Pi_{k+1}-\Pi_k$),
corresponding  to models  with an  effective temperature  as  close as
possible  to   the  value   of  $T_{\rm  eff}$   of  the   star  under
consideration.   

We first consider  the case of PG 1159-035.  According to its location
on  the $\log (T_{\rm  eff})-\log g$  plane, this  star should  have a
stellar  mass of  $\approx 0.536  M_{\odot}$. We refer this value as 
the ``spectroscopic mass'', $M_{\rm spec}$ (see Fig.  \ref{grave-obs}).
Our stability  analysis predicts  that a model  with this mass  at the
effective    temperature    of    PG   1159-035    is    pulsationally
unstable. However, this model should have a value of $\overline{\Delta
\Pi}_k \approx 23$  s, which is in   conflict with the observed
mean  value of  21.5 s.   To  have a  $\overline{\Delta \Pi_k}$  value
compatible with  the observed  period spacing the  stellar mass  of PG
1159-035 should be of $M_{\rm seis} \approx 0.558 M_{\odot}$. Thus, 
we found a discrepance of $\Delta M_*= M_{\rm seis} - M_{\rm spec} 
\approx 0.022 M_{\odot}$, somewhat lower than that reported in the 
literature ($M_{\rm spec} \approx 0.54 M_{\odot}$, Dreizler \& Heber 1998; 
$M_{\rm seis} \sim 0.59 M_{\odot}$, Kawaler \& Bradley 1994). We note that 
a model with $M_* \approx 0.558 M_{\odot}$ at
$140\,000$  K  should be  pulsationally  stable  in  the frame  of  our
stability analysis. 

In   the  case   of  RX   J2117+3412,  the   evolutionary   tracks  of
Fig. \ref{grave-obs} suggest a  stellar mass of about $0.72 M_{\odot}$
(see Miller  Bertolami \& Althaus  2006).  A model with this mass and at the 
$T_{\rm eff}$ of this star should be 
pulsationally unstable. The corresponding  value of
$\overline{\Delta \Pi_k}$ should  be lower  than $16$ s, clearly at odds
with  the measured  mean value  of  21.5 s.   So, in  order to  have
$\overline{\Delta \Pi_k}$ values comparable  to the observed, we should 
be forced to consider models with masses lower than $ 0.57 M_{\odot}$.
Thus, for RX J2117+3412 we found a large disagreement between the 
spectroscopic and the seismic mass ($\Delta M_* \approx -0.15 M_{\odot}$)
and in the opposite direction than for PG 1159-035. Note that models   
with  such  low  masses   (at  the  effective
temperature of  RX J2117+3412) are  outside of the  instability domain
(see Fig. \ref{grave-obs}).

Finally,  we  have  the cases  of  PG  2131+066,  PG 1707+427  and  PG
0122+200.   According to our  evolutionary tracks,  the mass  of these
stars should  be of  $\approx 0.55 M_{\odot}$  for PG 2131+066  and of
$\approx 0.53 M_{\odot}$ for PG  1707+427 and PG 0122+200.  Note that,
however,  in  order  to   have  values  of  $\overline{\Delta  \Pi}_k$
compatible with the observed  mean period spacings, the stellar masses
should  be substantially  larger,  of about  $0.58  M_{\odot}$ for  PG
2131+066 ($\Delta M_* \approx 0.03 M_{\odot}$), $0.57 M_{\odot}$  for PG 1707+427 ($\Delta M_* \approx 0.04 M_{\odot}$), and of  $ 0.65 M_{\odot}$
for PG 0122+200 ($\Delta M_* \approx 0.12 M_{\odot}$).   
Models with such  high stellar masses (at
the  effective  temperature  of  the stars  under  consideration)  are
pulsationally  unstables   (see  Fig.   \ref{grave-obs}).  
Given  the spectroscopic uncertainties  in the  determination of 
$\log  g$, these solutions  could  be  still  compatible  with those  
derived  from  our evolutionary tracks and stability analysis.

We conclude that the stellar masses of naked GW Vir stars as
predicted by  our evolutionary tracks  are generally $4-20 \%$ 
lower  than those suggested by the period spacing data.  By the contrary, 
for the PNNV RX J2117+3412 the evolutionary tracks predict a spectroscopic 
mass about $ 25 \%$ {\it higher} than the seismic derivation.   
Although our full evolutionary PG1159 models hint at generally agreement 
between the spectroscopic and seismic masses of pulsating PG1159 stars, 
persisting discrepancies could still be reflecting a problem in the stellar 
modelling during the pulsing AGB phase of progenitor stars, as noted by 
Werner \& Herwig (2006).

\section{Summary and conclusions}
\label{conclusions}

In this  paper we  re-examined the pulsational  stability  
properties  of  GW Vir  stars.   We  performed
extensive nonadiabatic computations on PG1159 evolutionary models with
stellar masses  ranging from $0.530$  to $0.741 M_{\odot}$.   For each
sequence of  models, we computed  the complete evolutionary  stages of
PG1159 progenitors starting from the Zero Age Main Sequence. Evolution
was pursued  through the thermally  pulsing AGB and  born-again (VLTP)
phases to the domain of the  PG1159 stars.  The employment of such full
evolutionary PG1159 models  constitutes a substantial improvement over
previous studies on GW Vir stars regarding the stellar modelling.

Numerous  detailed  investigations about  pulsating  PG1159 stars have  been
performed on the basis of  artificial stellar models.  In spite of the
fact  that the  significant  pulsation damping  and  driving occur  in
PG1159  envelope  stars, the  employment  of  such simplified  stellar
configurations appear not  well justified in the case  of these stars.
This  is  in   contrast  to  the  situation  of   their  more  evolved
counterparts, the white dwarf stars, for which their thermo-mechanical
structure has relaxed  to the correct one by  the time the pulsational 
instability domains are  reached. The main  goal of the  present work 
has  been to
assess to what  degree the conclusions arrived at  in previous studies
on  PG1159  stars change  when  realistic  stellar configurations  are
adopted.

Our study confirms the following results, already known from previous 
studies:

\begin{itemize}

\item $g$-modes in PG1159 models are  excited by the 
$\kappa$-mechanism  due to partial ionization of  carbon and oxygen.
No abundance gradients between the driving region and the stellar
surface  are  necessary  to  drive $g$-mode  pulsations  at  the correct 
effective temperatures and period ranges. 

\item There exists a well-defined  instability domain with  a blue  
edge which is  strongly dependent on  the stellar mass. 

\item Different surface helium abundances lead  to sizeable 
differences in the precise location  of the  theoretical blue edge of the  
instability domain.

\item  The instability domain splits into two 
separated regions, one of them at high luminosities characterized by long 
periods, and the other at low luminosities, corresponding to shorter periods,
as recently demonstrated by GAS05.

\item All  pulsating   PG1159  stars  lay  into  the  
predicted instability domain in  the $\log (T_{\rm eff})-\log g$  plane. 

\item There is a  very good agreement between  
the full period spectrum observed in GW Vir stars  
and  the theoretical ranges  of unstable periods.

\item  The pulsation periods of excited modes decrease  with decreasing
luminosity (increasing surface gravity), in line with the observational 
trend.

\end{itemize}

As for our new findings, we mention: 

\begin{itemize}

\item  There exists a red   edge  of   the   instability  domain   
at  the   high-luminosity (low-gravity)  regime. This red edge is 
mass-dependent.

\item The border of the  instability domains  in the 
$\log T_{\rm eff}-\log \Pi$ plane at  the high-luminosity, 
long-period regime is well delineated.   

\item The  pulsating PG1159 Longmore 4
is located at the very red edge of the instability strip at high
luminosities, a fact that could be reflecting the surprising behaviour
observed in the spectral type of  this star (Werner et  al. 1992). 

\item Some  non-variables occupying the  instability strip
have standard helium abundances and the presence of  them
between  pulsators can not be explained through the 
argument of Quirion et al. (2004). 

\item   The pulsating nature and also the range of observed periods of 
PG 1153-035 --- the prototype of the GW Vir class --- 
are naturally accounted for by pulsationally unstable PG1159 models with a 
stellar  mass of $\sim  0.53-0.54 M_{\odot}$.  

\end{itemize}

Finally, we found that generally  the seismic masses (as inferred from
the  period spacings)  are somewhat  different from  the spectroscopic
masses, although the disagreement for the PG 1159-035 star is somewhat
alleviated   according   to    our   calculations.    The   persisting
discrepancies  could be  attributed to  a  number of  factors. On  the
observational side,  possible systematics errors  in the spectroscopic
determination  of  $g$  and   $T_{\rm  eff}$,  and/or  errors  in  the
measurement of the period spacings  of pulsating PG1159 stars. On the
other hand, differences in the  microphysics or the previous evolution
may alter the location of the post-AGB tracks (Blo\"ecker 1995). In fact,
it  has been  argued by  Werner \&  Herwig (2006)  that  the evolution
during the  TP-AGB (concerning third  dredge up efficiency  and TP-AGB
lifetimes)  may  be  key  in  determining  the  location  of  post-AGB
tracks. However, in preliminary simulations we have found that neither
third dredge up efficiency nor TP-AGB lifetimes play an important role
in determining the location of post-AGB tracks. It remains to
be seen if other physical assumptions like the overshooting efficiency
during the core  helium burning stage and early  AGB (that also define
the structure of the C-O  core and are completely free parameters) may
be playing a role in the location of post-AGB tracks. We are currently
performing  simulations  of  full  stellar evolution  sequences  under
different assumptions to clarify these issues.


\begin{acknowledgements}

We wish to  thank our anonymous referee for  the constructive comments
and  suggestions that  greatly improved  the original  version  of the
paper.  This research was supported in part by the PIP 6521 grant 
from CONICET.

\end{acknowledgements}


\begin{thebibliography}{}

\bibitem{} Alexander, D. R., \& Ferguson, J. W. 1994, ApJ, 437, 879%
\bibitem{} Althaus, L. G., Serenelli, A. M., Panei, J. A., 
C\'orsico, A.  H., Garc\'{\i}a-Berro, E., \& Sc\'occola,  C. G.  2005,
A\&A, 435, 631%
\bibitem{} Bl\"ocker, T. 2001, Ap\&SS, 275, 1%
\bibitem{} Bond, H. E., \& Meakes, M. G. 1990, AJ, 100, 788%
\bibitem{} Bradley, P. A., \& Dziembowski, W. A. 1996, ApJ, 462, 376%
\bibitem{} Ciardullo, R., \& Bond, H. E. 1996, AJ, 111, 2332%
\bibitem{} C\'orsico, A. H., \& Althaus, L. G. 2005, A\&A, 439, L31%
\bibitem{} C\'orsico, A. H., \& Althaus, L. G. 2006, A\&A, in press
({\tt astro-ph/0603736})%
\bibitem{} Cox, A. N. 2003, ApJ, 585, 975%
\bibitem{} Dreizler, S., \& Heber, U. 1998, A\&A, 334, 618%
\bibitem{} Fujimoto, M. Y. 1977, PASJ, 29, 331%
\bibitem{} Gautschy, A. 1997, A\&A, 320, 811%
\bibitem{} Gautschy, A., Althaus,  L. G., \& Saio, H. 2005, A\&A, 
438, 1013 (GAS05)%
\bibitem{} Grauer, A. D., Bond, H. E., Liebert, J., Fleming, T. A., 
\& Green, R. F.\ 1987, ApJ, 323, 271%
\bibitem{} Herwig, F. 2000, A\&A, 360, 952%
\bibitem{} Iglesias, C. A., \& Rogers, F. J. 1996, ApJ, 464, 943%
\bibitem{} Kawaler, S. D., et al. 2004, A\&A, 428, 969%
\bibitem{} Kawaler, S. D., et al. 1995, ApJ, 450, 350%
\bibitem{} Kawaler, S. D., \& Bradley, P. A. 1994, ApJ, 427, 415%
\bibitem{} Kawaler, S. D. 1988, ApJ, 334, 220%
\bibitem{} Kawaler, S. D., Winget, D. E., Hansen, C. J., 
\& Iben, I. 1986, ApJL, 306, L41%
\bibitem{} Koesterke, L., Dreizler, S., \& Rauch, T. 1998, A\&A, 330, 1041%
\bibitem{} Lee, U., \& Bradley, P. A. 1993, ApJ, 418, 855 %
\bibitem{} Miksa, S., Deetjen, J. L., Dreizler, S., et al. 2002, A\&A, 
389, 935%
\bibitem{} Miller Bertolami, M. M., \& Althaus, L. G. 2006, A\&A, 
in press ({\tt astro-ph/0603846})%
\bibitem{} Nagel, T., \& Werner, K.\ 2004, A\&A, 426, L45%
\bibitem{} O'Brien, M. S. 2000, ApJ, 532, 1078%
\bibitem{} O'Brien, M. S., et al. 1998, ApJ, 495, 458%
\bibitem{} Quirion, P. O., Fontaine, G., \& Brassard, P. 2004, ApJ, 610, 436%
\bibitem{} Reed, M. D., Kawaler, S. D., \& O'Brien, M. S. 2000, ApJ, 545, 429%
\bibitem{} Saio, H. 1996, ASP Conf. Ser.96: Hydrogen Deficient Stars, 
96, 361%
\bibitem{} Sch\"onberner, D. 1979, A\&A, 79, 108%
\bibitem{} Silvotti, R., Dreizler, S., Handler, G., \& Jiang, X. J. 1999, 
A\&A, 342, 745%
\bibitem{} Starrfield, S. G., Cox, A. N., Hodson, S. W., \& Pesnell, 
W. D. 1983, ApJ, 268, L27%
\bibitem{} Tassoul, M., Fontaine, G., \& Winget, D. E. 
1990, ApJS, 72, 335%
\bibitem{} Unglaub, K., \& Bues, I. 2000, A\&A, 359, 1042%
\bibitem{} Unno, W., Osaki, Y., Ando, H., Saio, H., \& Shibahashi, H.  
1989,  Nonradial Oscillations  of  Stars, University  of Tokyo  Press,
2nd. edition%
\bibitem{} Vauclair, G., Solheim, J. E., \& {\O}stensen, R. H. 2005, 
A\&A, 433, 1097%
\bibitem{} Vauclair, G., et al. 2002, A\&A, 381, 122%
\bibitem{} Werner, K., \& Herwig, F. 2006, PASP, 118, 183%
\bibitem{} Werner, K., Hammann, W.-R., Heber, U., Napiwotzki, R., 
Rauch, T., \& Wessolowski, U. 1992, A\&A, 259, L69%
\bibitem{} Winget, D. E., et al. 1991, ApJ, 378, 326%

\end{thebibliography}
\end{document}